\documentclass[prl,twocolumn,superscriptaddress,longbibliography]{revtex4-2}
\usepackage{amssymb,amsmath,amsfonts}
\usepackage{graphicx}
\usepackage{color,xcolor}
\usepackage{hyperref}
\usepackage{bm}
\usepackage{multirow}
\usepackage{natbib}
\usepackage{comment}
\usepackage{caption}
\usepackage{placeins}

\begin{document}

\title{
Onset of Melting in Finite Ion Crystals: The Role of Structural Isomerization
}

\author{B. V. Pashinsky}

\affiliation{Department of Physics, University of Washington, Seattle, Washington 98195, USA}

\author{B. Blinov}

\affiliation{Department of Physics, University of Washington, Seattle, Washington 98195, USA}

\date{\today}

\begin{abstract}
We theoretically investigate the microscopic mechanism of melting in finite two-dimensional ion crystals confined in anisotropic traps. Building on our previous studies of melting probability as a function of temperature and anisotropy, we extend the analysis to crystal sizes ranging from $N=4$ to $N=101$ ions and examine how structural isomerization modifies the melting pathways. Using molecular dynamics simulations together with Metropolis-Hastings equilibrium sampling, we analyze radial fluctuations, angular disorder, Lindemann parameters, and isomer-dependent energy landscapes in these Coulomb crystals. We find that crystals with identical particle number and trap anisotropy can exhibit substantially different melting behavior depending on the metastable structural isomer they occupy. Contrary to a simple soft-mode instability picture, the observed melting process is characterized by a gradual thermally-driven accumulation of radial and angular disorder. Structural isomerization reshapes the fluctuation channels through which delocalization develops, leading to nonuniform melting probabilities across temperature and anisotropy space. These results provide a unified microscopic picture of melting in finite ion crystals and clarify the role of structural rearrangements in finite-size phase transitions.
\end{abstract}

\maketitle

\section{Introduction}

Finite-size trapped ion crystals provide a highly controllable setting for studying strongly interacting mesoscopic systems. Their equilibrium structures and dynamics are governed by the interplay of long-range Coulomb repulsion, external confinement, and thermal fluctuations, giving rise to a rich variety of structural configurations,
collective phenomena, and phase transitions ~\cite{Schiffer1993PhaseTransitions,Dubin1993StructuralTransitions,Enzer2000,Fishman2008,Lemmer2015,Mizukami2025}.
In addition to their fundamental interest as finite Coulomb systems, trapped-ion
crystals are relevant for quantum information processing ~\cite{Haffner2008,BrownKimMonroe2016}, programmable quantum simulation
~\cite{BlattRoos2012,Monroe2021,FossFeig2025},
and precision sensing ~\cite{Gilmore2021,Ludlow2015,Morigi2026}.
Two-dimensional ion crystals are especially attractive in this context because they provide an experimentally accessible route toward extended interacting systems beyond one-dimensional ion chains ~\cite{DOnofrio2021,Ivory2020,Kato2022,Kiesenhofer2023,Duca2023,Qiao2024}.

The thermodynamics of finite Coulomb crystals differs qualitatively from that of bulk crystals. Since the number of particles is finite, long-range translational order is replaced by a shell structure, boundary effects, and a discrete set of metastable configurations. Consequently, melting in such systems need not occur as a sharp, universal phase transition. Instead, disorder may appear through shell-dependent motion, inter-shell rearrangements, local defect formation, or orientational delocalization~\cite{Bedanov1994,Lozovik1998,Kong2003,Yurtsever2005,Apolinario2006Inhomogeneous,Duca2023}. This makes finite ion crystals a useful platform for studying how microscopic structure controls the onset of melting.

A useful point of comparison is provided by confinement-driven structural transitions in trapped-ion systems. In linear ion chains, the transition to a zigzag configuration is associated with the softening of a collective transverse mode and can be described using a normal-mode or Landau-type picture~\cite{Fishman2008}. This suggests a natural question for finite two-dimensional crystals: is melting also governed by the growth of fluctuations along a single soft collective coordinate, or does it proceed through a more distributed loss of localization involving several shell-dependent degrees of freedom? In contrast to the zigzag transition, melting is not merely a change between two nearby equilibrium structures; it involves thermal motion, finite energy barriers, and the loss of particle localization over time.

An additional complication is structural isomerization. Finite Coulomb crystals can possess multiple local minima with the same number of particles but different shell occupations and geometries. Such isomers are separated by finite barriers and may have different local energy landscapes, normal modes, and susceptibilities to thermal fluctuations. The role of metastable structures and finite-size effects has been studied in classical two-dimensional Coulomb clusters~\cite{Yurtsever2005}, and recent work has used two-dimensional Coulomb crystals to emulate isomerization dynamics in a controlled few-ion system~\cite{Mizukami2025}. These studies motivate treating isomeric structure not as a secondary detail, but as a possible control parameter for melting pathways.

In our previous work, we studied structural transitions and melting of two-dimensional ion crystals in a radio-frequency trap and mapped the melting probability as a function of temperature and trap anisotropy~\cite{Pashinsky2025}. That work established the basic phenomenology: melting temperature depends strongly on ion number and anisotropy, and the outer shell typically loses localization before the inner shell. However, the microscopic onset of melting and the role of structural isomerization remained unresolved. In particular, irregularities in melting probability maps suggest that crystals with the same ion number and similar trap parameters may nevertheless follow different disordering pathways.

%

Here, we investigate the onset of melting in finite two-dimensional ion crystals, with emphasis on the role of structural isomerization. We extend the numerical analysis to crystals containing (N=4) to (N=101) ions, combine molecular-dynamics simulations with Metropolis-Hastings equilibrium sampling~\cite{Metropolis1953,Hastings1970}, and analyze shell-resolved observables including radial fluctuations, angular disorder, and Lindemann parameters. We find that melting is predominantly associated with the progressive disordering of the outer shell and ultimately manifests as angular delocalization along the shell. Although low-frequency collective modes shape the local energy landscape, our results do not support a universal one-coordinate soft-mode instability as the melting onset mechanism. Instead, melting appears as a gradual, thermally driven loss of localization whose pathway is modified by the underlying isomeric structure of the crystal.

\section{Physical System}

\subsection{Experimental System}

The experimental platform considered in this work is identical to that described in Ref.~\cite{Pashinsky2025}. Briefly, laser-cooled $^{138}\mathrm{Ba}^{+}$ ions are confined in a radio-frequency Paul trap specifically designed for the formation of radial two-dimensional Coulomb crystals~\cite{Ivory2020,Kato2022}. Doppler cooling is performed using the $6S_{1/2}\rightarrow6P_{1/2}$ transition at 493~nm together with the 650~nm repumping transition. Ion fluorescence is collected by a high numerical-aperture parabolic mirror and imaged onto an EMCCD camera, allowing individual ions to be spatially resolved.

The present work is primarily theoretical. Experimental observations are used as qualitative motivation and for comparison with the numerical results presented below. Since the experimental setup has been described in detail in ~\cite{Pashinsky2025}, only the aspects directly relevant to the numerical model are summarized here.

\subsection{Effective Model}

Within the pseudopotential approximation, the potential energy of a crystal containing $N$ ions is given by:

\begin{equation}
U=
\frac{m}{2}
\sum_{i=1}^{N}
\left(
\omega_y^2y_i^2+
\omega_z^2z_i^2
\right)
+
\frac{e^2}{4\pi\varepsilon_0}
\sum_{i<j}
\frac{1}{r_{ij}},
\label{eq:potential}
\end{equation}

\noindent where $m$ and $e$ are the ion mass and charge, respectively, $\omega_y$ and $\omega_z$ are the secular trapping frequencies along the $y$ and $z$ principal trap axes, respectively, and $r_{ij}$ denotes the distance between ions $i$ and $j$. Throughout this work, only the in-plane equilibrium configurations are considered.

The geometry of the crystal is controlled by the trap anisotropy 
$\gamma={\omega_z}/{\omega_y}$.
Varying $\gamma$ changes the equilibrium shell structure of the crystal and produces a sequence of structural transitions between different stable configurations.

For a given pair of external parameters $(N,\gamma)$ the potential-energy landscape may possess multiple local minima. Each minimum corresponds to a mechanically stable equilibrium configuration containing the same number of ions but differing in shell occupation and ion arrangement. Throughout this work these metastable equilibrium configurations are referred to as \emph{structural isomers}. Different isomers are separated by finite potential-energy barriers and generally exhibit different normal-mode spectra, local stiffness, and thermal stability~\cite{Yurtsever2005,Mizukami2025}.

Unlike infinite crystals, where the equilibrium structure is uniquely determined by the external parameters, finite Coulomb crystals therefore possess an additional structural degree of freedom associated with isomerization. One of the central goals of the present work is to investigate how this additional complexity modifies the microscopic pathways leading to melting.

The equilibrium configurations analyzed in this work were obtained by numerical minimization of Eq.~(\ref{eq:potential}) for ion numbers ranging from $N=4$ to $N=101$. These equilibrium structures provide the starting point for all subsequent molecular-dynamics simulations, normal-mode calculations, and thermal analyses presented in the following sections.

\section{Numerical Approach}

For each ion number $N$ and trap anisotropy $\gamma$, we obtained
equilibrium crystal configurations by minimizing the potential energy
defined above. The thermal evolution of these configurations was then
investigated using classical molecular-dynamics simulations. Independent
trajectories were initialized with randomly generated velocities
corresponding to a prescribed temperature $T$.

The melting probability $P_{\mathrm{melt}}(N,\gamma,T)$ was estimated
from ensembles of trajectories using an operational criterion for the
loss of particle localization. To investigate the microscopic melting
dynamics, we also analyzed normal modes, outer-shell radial and angular
motion, and the time evolution of the Lindemann parameter. Equilibrium
Monte Carlo simulations provided an independent comparison through
heat-capacity calculations. Detailed numerical procedures, definitions
of the observables, and simulation parameters are provided in the
 \hyperref[sec:supplementary_information]{``Supplementary Information" section}.

\onecolumngrid

\begin{center}
\includegraphics[
    height=0.82\textheight,
    keepaspectratio
]{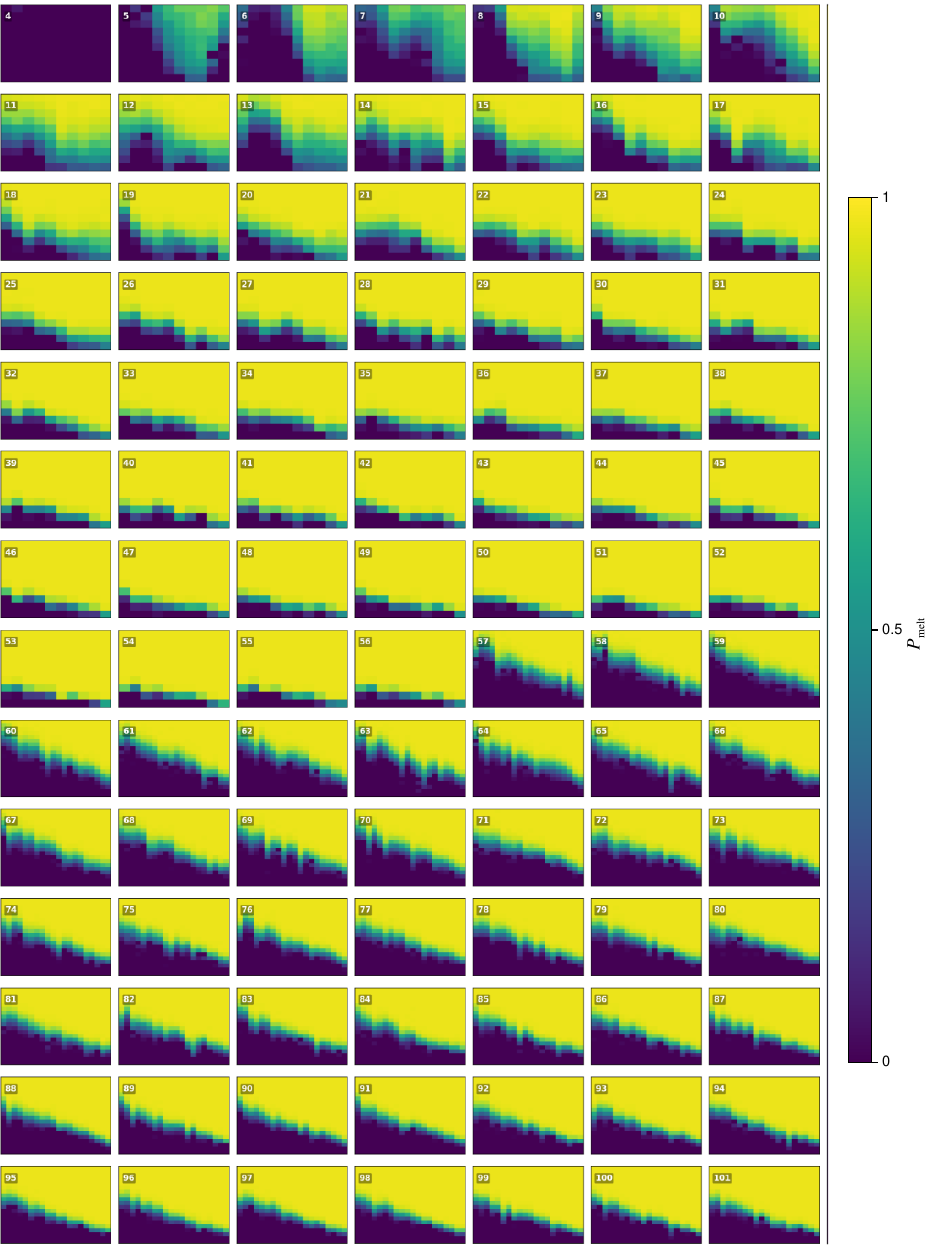}
\captionsetup{hypcap=false}
\captionof{figure}{
Melting phase diagrams of finite two-dimensional Coulomb crystals containing $4 \le N \le 101$ ions. Each panel shows the melting probability $P_{\mathrm{melt}}(T,\gamma)$ obtained from molecular-dynamics simulations for a fixed particle number $N$. The trap anisotropy varies in the range $0.5 \le \gamma \le 1$, where $\gamma=1$ corresponds to the isotropic trap.
For clarity, different temperature ranges are used for the two groups of particle numbers. For particle numbers $4\leq N\leq56$, the melting probabilities were evaluated at 10 uniformly spaced temperatures over the interval $0.01\leq T\leq0.10$. For $57\leq N\leq101$, the lower melting temperatures required a narrower and more finely resolved temperature grid consisting of 20 points over $0\leq T\leq0.04$. The $N=56$ crystal was simulated using both temperature grids for comparison; the wider interval is shown for consistency with the lower-$N$ group. The color scale represents the melting probability from 0 (dark blue) to 1 (yellow).}
\label{fig:Atlas}

\end{center}

\twocolumngrid

\section{Results}
\subsection{Global melting phase diagrams}

Using the numerical workflow described in \hyperref[sec:molecular_dynamics]{Molecular dynamics subsection}, we systematically investigated the melting of finite two-dimensional Coulomb crystals containing between 4 and 101 ions. For each crystal size, molecular-dynamics simulations were performed over a range of temperatures and trap anisotropies, yielding the melting phase diagrams shown in Fig.~\ref{fig:Atlas}.

To our knowledge, this is the first comprehensive survey of melting phase diagrams covering the entire family of stable finite two-dimensional Coulomb crystals. The resulting dataset enables the evolution of melting behavior to be followed continuously as a function of crystal size.

Several general trends are immediately apparent. First, the melting temperature depends strongly on the trap anisotropy. Second, the character of the melting transition changes systematically with increasing particle number. While the detailed melting boundaries become increasingly structured with particle number, their normalized temperature width decreases on average, as quantified below.

Despite these global trends, neighboring crystal sizes often exhibit markedly different melting behavior. As will be shown below, these differences originate from changes in the equilibrium shell structure and from the existence of multiple metastable isomers.

\subsection{Size dependence of the melting transition}

Strong size dependence, broadened transitions, and structure-specific
irregularities are common features of finite clusters
\cite{Robbins1990FiniteSize,Baletto2005Nanoclusters}.

The global phase diagrams in Fig.~\ref{fig:Atlas} suggest that the melting transition becomes progressively narrower as the number of ions increases. To quantify this behavior, we define the normalized transition width as

\begin{equation}
W(N)
=
\frac{T_{90\%}-T_{10\%}}{T_{50\%}},
\label{eq:melting_width}
\end{equation}

\noindent where $T_{p}$ is the temperature at which the melting probability reaches value $p$. Thus, $T_{10\%}$ and $T_{90\%}$ delimit the main crossover
region, while normalization by $T_{50\%}$ removes the overall shift of the melting temperature with crystal size.

\begin{figure}[t]
\includegraphics[width=\columnwidth]{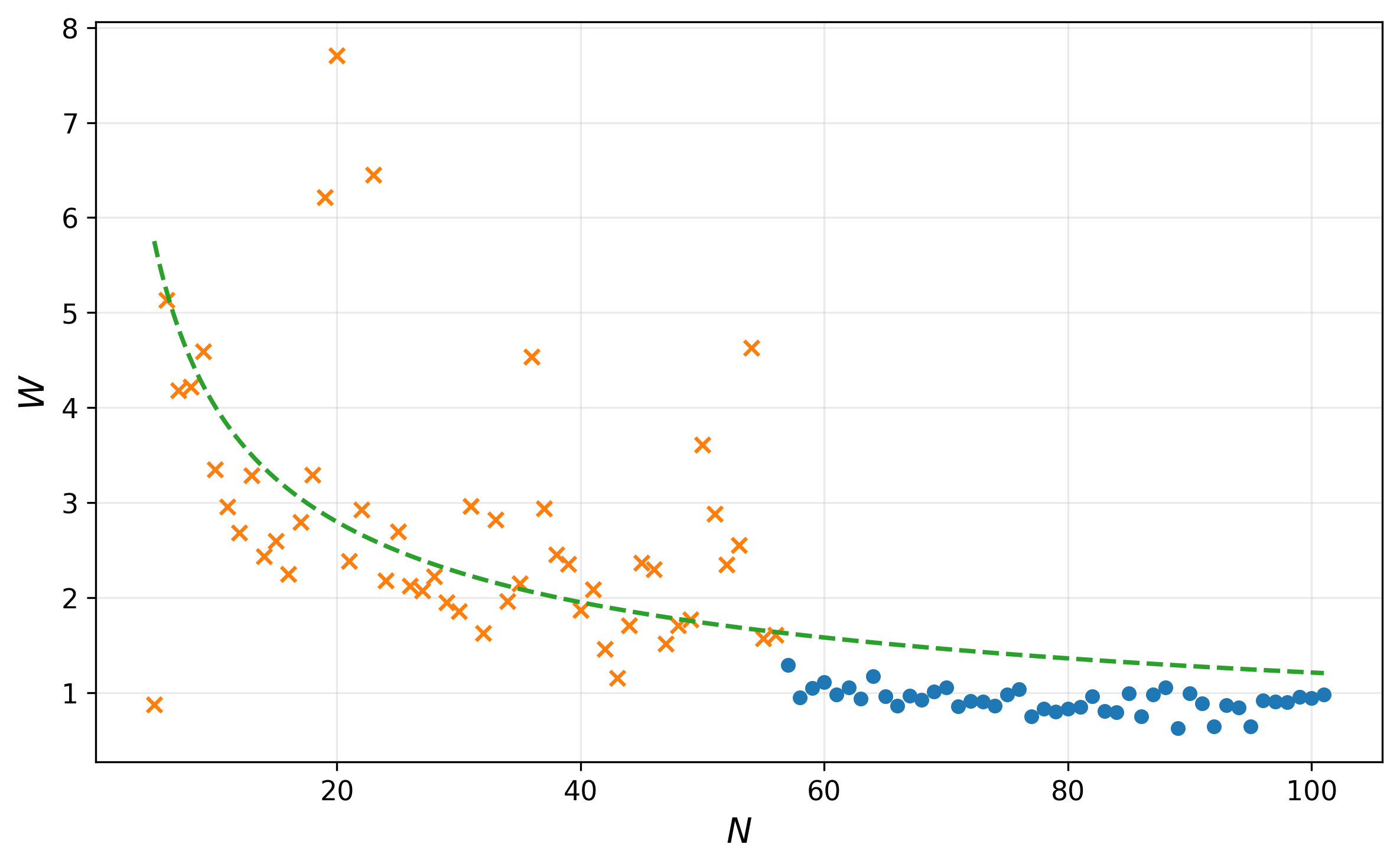}
\caption{
Normalized melting-transition width
$W=(T_{90}-T_{10})/T_{50}$ as a function of ion number $N$ at $\gamma=1$.
Here, $T_p$ denotes the temperature at which the melting probability reaches
$P_{\mathrm{melt}}=p$. Filled circles indicate cases in which all three threshold temperatures $T_{10}$, $T_{50}$, and $T_{90}$ lie within the simulated temperature range, whereas crosses indicate cases in which at least one threshold temperature was obtained by linear extrapolation. The dashed curve shows the power-law dependence $W\propto N^{-0.52}$. For $N\leq56$, the melting probabilities were evaluated at 10 uniformly spaced temperatures over $0.01\leq T\leq0.10$, whereas for $N\geq57$, 20 temperatures over $0\leq T\leq0.04$ were used.\label{fig:Width}
}

\end{figure}

Figure~\ref{fig:Width} shows $W(N)$ for crystals containing
$4\leq N\leq101$ ions at $\gamma=1$. Despite substantial fluctuations between
neighboring particle numbers, the overall transition width decreases with
increasing $N$. A power-law fit gives

\begin{equation}
W(N)\propto N^{-0.52},
\end{equation}

\noindent which matches very closely the inverse-square-root dependence. This result indicates that
the melting crossover becomes sharper as the finite crystal grows, consistent
with the emergence of increasingly collective behavior in larger systems.

The deviations from a smooth size dependence are not purely statistical.
Neighboring crystal sizes may have different shell occupations and equilibrium
geometries, which modify their thermal stability and broaden or narrow the
melting crossover. These finite-size structural effects produce the pronounced
scatter around the average trend and motivate the isomer-resolved analysis
presented below.

For some crystal sizes, one or more of the probability levels
$P_{\rm melt}=0.1$, $0.5$, and $0.9$ lie outside the simulated temperature
interval. In these cases, the corresponding crossing temperatures were
estimated by linear extrapolation of the monotonic melting-probability curve.
The temperature grid also changes between
$N=56$ and $N=57$, because the lower melting temperatures of the larger
crystals required a narrower temperature interval to resolve the transition.

\subsection{Structural isomerization}
\label{sec:structural_isomerization}

The melting phase diagrams discussed above describe only the probability of melting as a function of the external control parameters. However, they do not uniquely determine the microscopic structure of the crystal. For finite Coulomb systems, the potential-energy landscape generally contains several local minima corresponding to distinct equilibrium configurations with the same ion number and trap anisotropy. Throughout this work these metastable configurations are referred to as structural isomers. Multiple metastable structures and thermally activated rearrangements are
well-known features of finite clusters \cite{Yurtsever2005,Mizukami2025,Ayyadevara2026ThermalActivation}.

\begin{figure*}

\includegraphics[width=\textwidth]{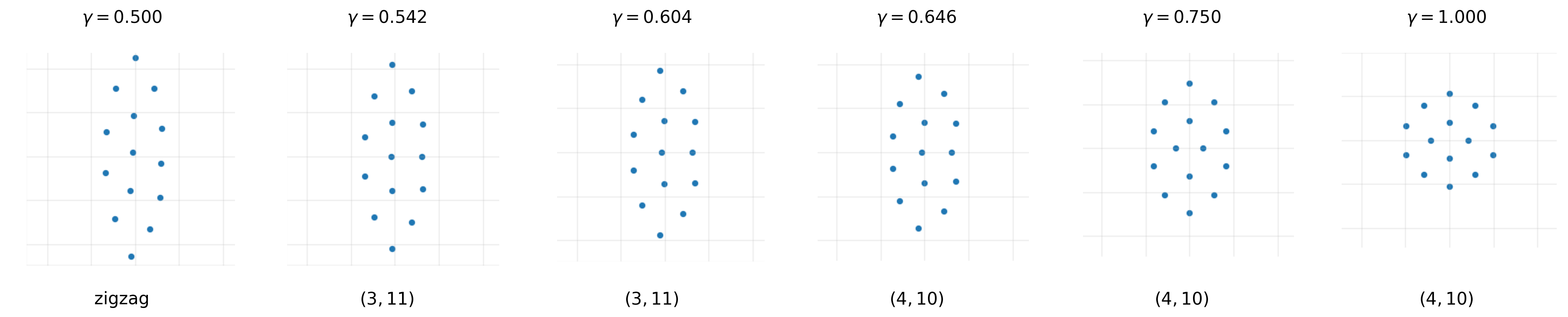}
\caption{Evolution of equilibrium configurations of a 14-ion crystal as the trap anisotropy is varied.
The crystal undergoes a sequence of structural transitions from the zigzag configuration to the two-shell isomers $(3,11)$ and $(4,10)$.
Within the coexistence interval both two-shell configurations are mechanically stable, corresponding to distinct local minima of the potential-energy landscape.
\label{fig:isomer14}}
\end{figure*}

Figure~\ref{fig:isomer14} illustrates the evolution of the equilibrium structure for a 14-ion crystal as the trap anisotropy is varied. As the crystal approaches the structural transition, two different shell occupations, $(3,11)$ and $(4,10)$, where the numbers in the brackets represent the shell occupancy starting from the innermost shell, become mechanically stable over a finite interval of anisotropy. Which structure is obtained depends on the initial conditions used during energy minimization, indicating that both correspond to local minima of the potential-energy landscape separated by a finite energy barrier.

\begin{figure*}

\includegraphics[width=\textwidth]{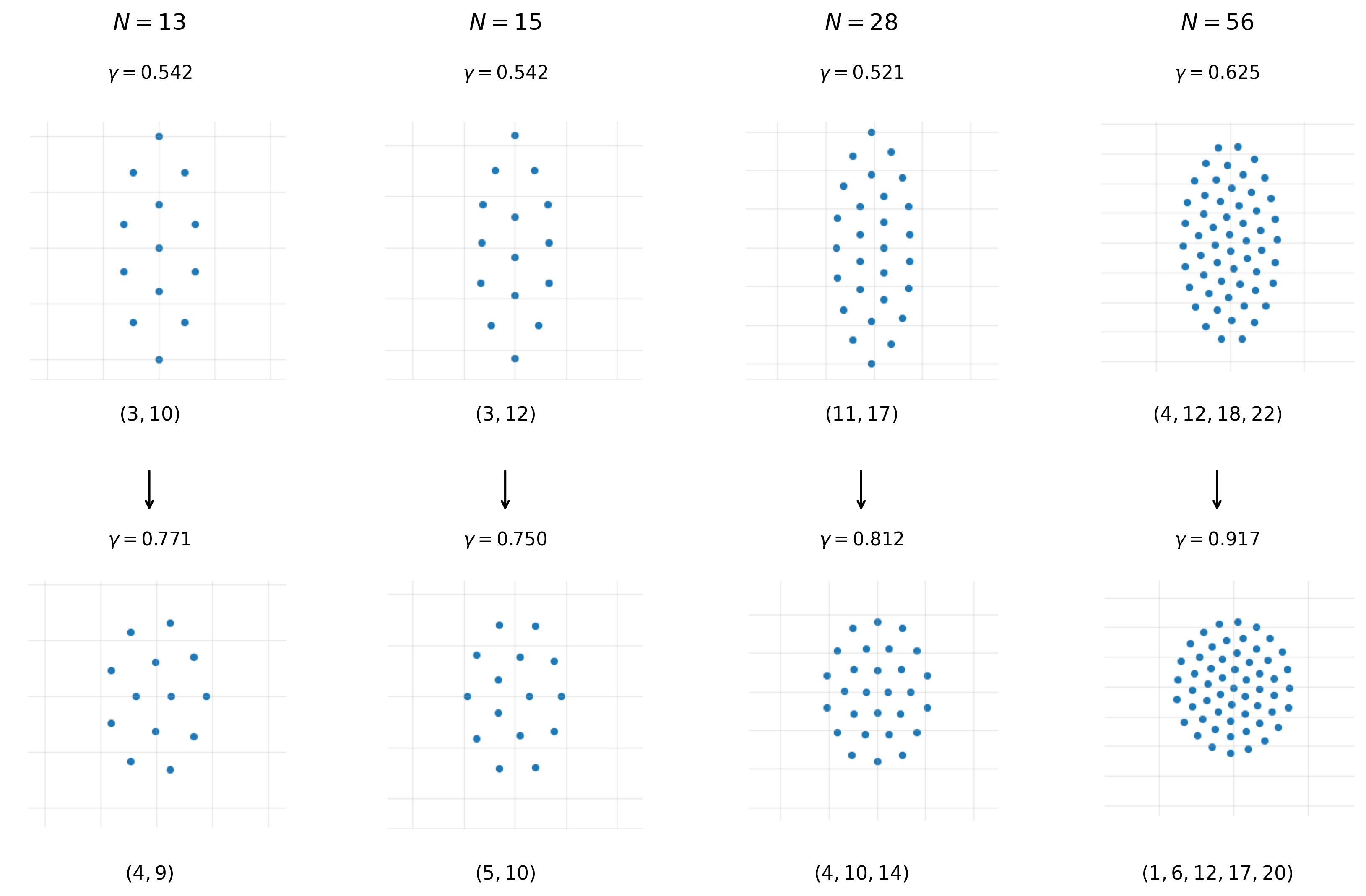}
\caption{Representative examples of structural isomerization for several crystal sizes.
For identical external control parameters, more than one mechanically stable equilibrium configuration can exist.
The coexistence of multiple structural isomers is therefore a generic property of finite two-dimensional Coulomb crystals.
\label{fig:isomer_examples}}
\end{figure*}

Structural isomerization is not unique to the 14-ion crystal, of course. Representative examples for several other crystal sizes are shown in Fig.~\ref{fig:isomer_examples}. Similar coexistence of multiple equilibrium configurations is observed over a broad range of particle numbers, demonstrating that metastability is a generic feature of finite two-dimensional Coulomb crystals rather than a peculiarity of a single system.

The existence of multiple metastable structures naturally raises the question of whether different isomers exhibit identical thermal behavior. Since each minimum possesses its own local curvature and normal-mode spectrum, different isomers may respond differently to thermal fluctuations even under identical external conditions. This question is addressed in the following section.

\subsection{Isomer-dependent melting}
\label{sec:isomer_dependent_melting}

Multiple structural isomers may coexist for the same number of ions and trap
anisotropy. Thermal fluctuations can induce transitions between these
metastable configurations, while differences in their equilibrium geometry,
local curvature, and normal-mode spectra can result in different thermal
stability. 
Consequently, individual isomers may follow distinct melting pathways and exhibit different melting probabilities even under identical external conditions. This behavior is consistent with the broader cluster literature, where structural, thermodynamic, and kinetic effects are strongly coupled and metastable configurations can substantially modify the melting scenario \cite{Baletto2005Nanoclusters,Yurtsever2005,Ferreira2005}.


As a representative example, we consider the $N=14$ crystal at
$\gamma=0.63$, where two competing isomers, $(3,11)$ and $(4,10)$, coexist.
Independent molecular-dynamics simulations were performed starting from each
equilibrium configuration. For every trajectory we recorded both the initial
isomer and the occurrence of melting, allowing us to evaluate the conditional
melting probability for each structural state.

\begin{figure}
    \centering
    \includegraphics[width=0.45\textwidth]{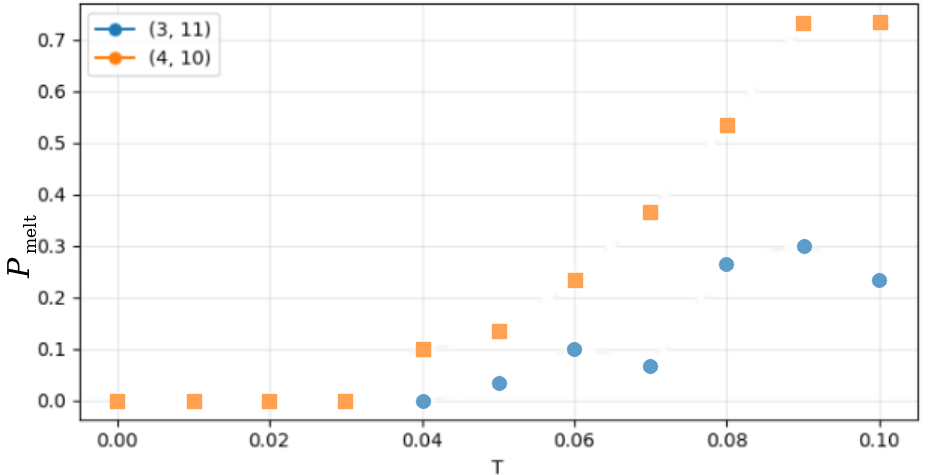}
    \caption{
    Conditional melting probability for the two competing $N=14$ isomers at
    $\gamma=0.63$. The probability is evaluated separately for trajectories
    initialized in the $(3,11)$ and $(4,10)$ configurations. The two isomers
    exhibit substantially different thermal stability despite having the same
    particle number and trap anisotropy.
    }
    \label{fig:conditional_melting}
\end{figure}

Figure~\ref{fig:conditional_melting} demonstrates a pronounced dependence of the melting probability on the initial structural state. Over the investigated temperature range, the $(4,10)$ configuration exhibits a substantially higher melting probability than the competing $(3,11)$ isomer. The difference becomes particularly pronounced at intermediate temperatures, demonstrating that the particular metastable structure occupied by the crystal substantially modifies its thermal stability.

To investigate the microscopic origin of this difference, we next examine the lowest-frequency normal modes of the two equilibrium structures. We characterize each mode by two quantities introduced in the normal-mode analysis:
the outer-shell participation $P_{\mathrm{out}}$, which measures the fraction of the total mode amplitude carried by the outer-shell ions, and the tangential fraction $f_{\theta}$, which measures the fraction of the total mode
amplitude corresponding specifically to tangential motion of the outer shell.

\begin{table}[t]
\caption{
Character of the lowest-frequency positive normal mode for the competing $N=14$ isomers at $\gamma=0.63$. $P_{\mathrm{out}}$ denotes the fraction of the mode amplitude localized on the outer shell, while $f_{\theta}$ denotes the fraction corresponding to tangential motion of the outer-shell ions.
The last column gives the effective small-angle stiffness
$k_\theta$ for rigid rotation of the outer shell.
}
\label{tab:isomer_modes}
\begin{ruledtabular}
\begin{tabular}{cccc}
Isomer &
$P_{\mathrm{out}}$ &
$f_{\theta}$ &
$k_{\theta}$  \\
\hline
$(3,11)$ & 0.959 & 0.895 & 6.30  \\
$(4,10)$ & 0.999 & 0.916 & 8.03 
\end{tabular}
\end{ruledtabular}
\end{table}

As shown in Table~\ref{tab:isomer_modes}, the lowest-frequency mode is strongly localized on the outer shell for both isomers. Approximately $96\%$ of the mode amplitude resides on the outer shell for the $(3,11)$ configuration, while the corresponding value is essentially $100\%$ for $(4,10)$. Moreover, the mode is predominantly tangential: approximately $90\%$ and $92\%$ of the total mode amplitude corresponds to tangential outer-shell motion for the $(3,11)$ and $(4,10)$ configurations, respectively.

Thus, the geometrical character of the softest collective excitation is similar in the two isomers: in both cases it predominantly corresponds to azimuthal motion of the outer shell. Their energetic response along this collective coordinate, however, is different.

The relative rotation of concentric shells provides a natural collective coordinate for finite Coulomb crystals. Recent calculations for ion Coulomb crystals have shown that the corresponding rotational energy landscape can depend strongly on shell geometry, ion number, and confinement \cite{Rueffert2026}. We therefore characterize the two competing isomers by rigidly rotating the outer shell relative to the inner shell and evaluating the resulting change in potential energy,

\begin{equation}
    \Delta E(\delta)
    =
    E(\delta)-E(0),
\end{equation}

\begin{figure}
    \centering
    \includegraphics[width=0.45\textwidth]{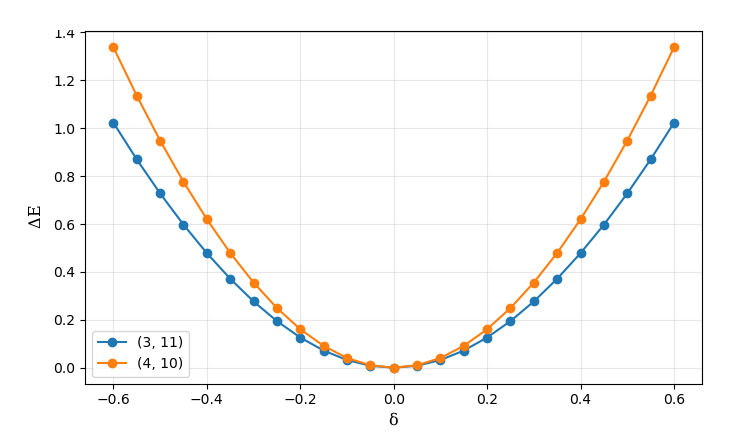}
    \caption{
    Change in potential energy produced by a rigid rotation of the outer shell
    for the competing $N=14$ isomers at $\gamma=0.63$. The $(4,10)$
    configuration exhibits a steeper energy increase than the $(3,11)$
    configuration, indicating stronger angular confinement. The corresponding
    small-angle stiffnesses are $k_{\theta}\simeq8.03$ and
    $k_{\theta}\simeq6.30$, respectively.
    }
    \label{fig:rotational_barrier}
\end{figure}

\noindent The resulting energy profiles $E(\delta)$ are shown in
Fig.~\ref{fig:rotational_barrier}, where $\delta$ is the rigid angular displacement of the outer shell from its equilibrium orientation. The two curves correspond to the $(3,11)$ and $(4,10)$ isomers.

Near equilibrium, the rotational energy can be approximated by

\begin{equation}
    \Delta E(\delta)
    \simeq
    \frac{1}{2}k_{\theta}\delta^2,
\end{equation}

\noindent which defines the effective angular stiffness $k_{\theta}$. The $(4,10)$ configuration has a larger stiffness,
$k_{\theta}\simeq8.03$, compared with $k_{\theta}\simeq6.30$ for $(3,11)$.

Interestingly, the $(4,10)$ configuration exhibits the larger effective rotational stiffness and the larger energy cost for rigid outer-shell rotation, despite showing the higher melting probability at a given temperature in Fig.~\ref{fig:conditional_melting}. Thus, the different thermal stability of the two isomers cannot be understood solely from the stiffness of a single collective rotational coordinate.

Together with the predominantly tangential character of the lowest-frequency modes, this observation motivates a direct analysis of the full molecular-dynamics trajectories. In the following section, we therefore determine the temporal ordering of radial fluctuations, loss of localization, and large-amplitude tangential motion during melting.

\subsection{Melting dynamics}
\label{sec:melting_dynamics}

The phase diagrams presented above determine the range of external parameters for which melting occurs, but they do not reveal the microscopic sequence of
events leading to the loss of crystalline order. To identify this sequence, we analyzed both individual molecular-dynamics trajectories and ensemble-averaged
observables characterizing radial motion, angular disorder, and particle delocalization.

A natural initial hypothesis is that melting is driven directly by excitation of the lowest-frequency collective mode. For the configurations considered
here, this mode is strongly localized on the outer shell and has a predominantly tangential character. Such a mode provides a natural coordinate for describing
azimuthal motion of the outer ions. If melting were caused by a soft-mode instability, one would therefore expect large tangential motion to precede or coincide with the loss of crystalline localization.

\begin{figure}
    \centering
    \includegraphics[width=\columnwidth]{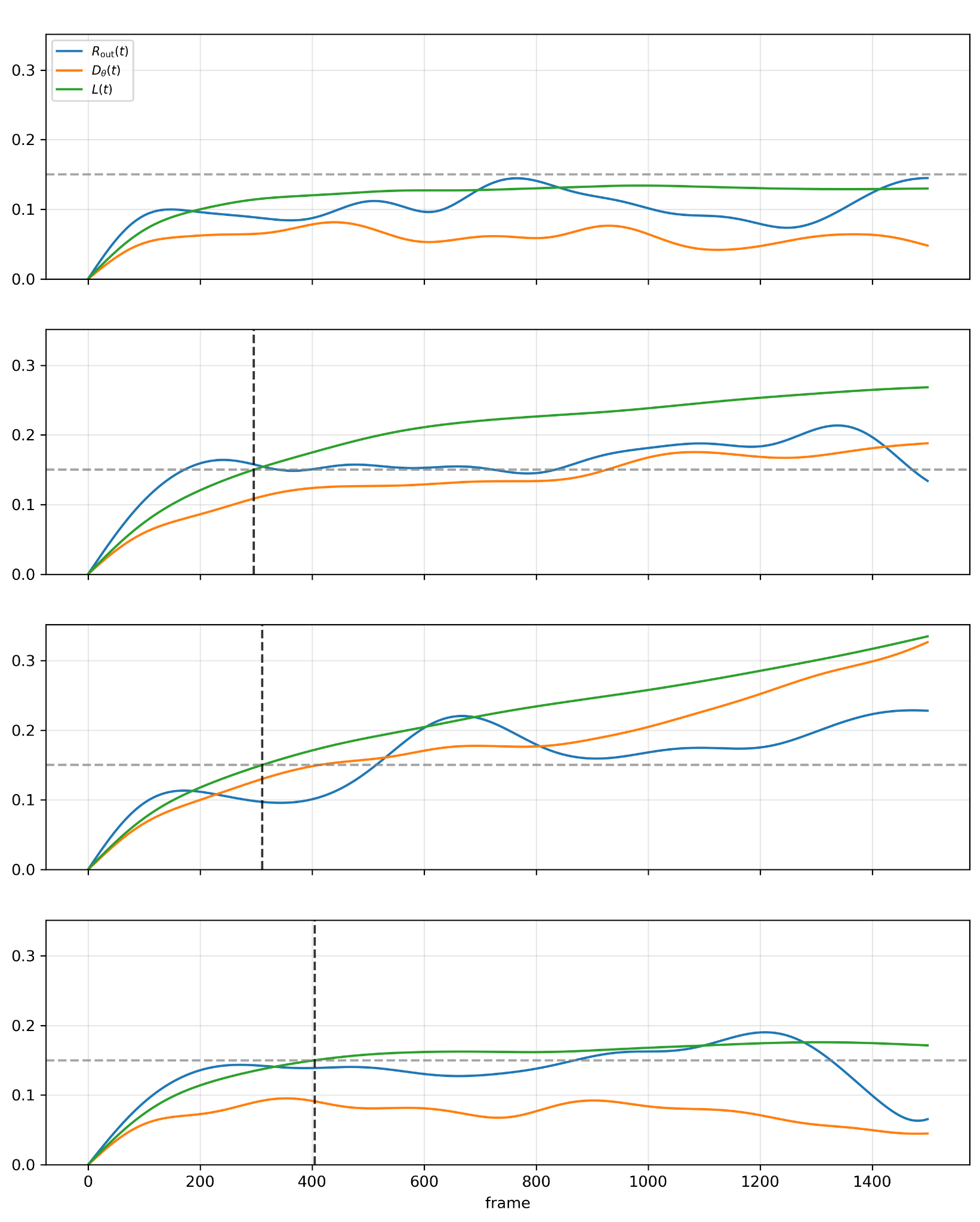}
    \caption{
    Representative time evolution of the outer-shell radial fluctuation $R_{\mathrm{out}}(t)$, angular-disorder parameter $D_{\theta}(t)$, and Lindemann parameter $L(t)$ for several molecular-dynamics trajectories plotted as a function of computational time ("frame").
    The vertical dashed line indicates the melting time, and the horizontal dashed line marks the Lindemann threshold. The first three trajectories melt, whereas the fourth remains localized during the simulated interval. Pronounced radial excursions may occur before melting and may also occur without subsequent melting.
    }
    \label{fig:single_trajectories}
\end{figure}

Figure~\ref{fig:single_trajectories} shows representative time traces obtained from the individual trajectories of ions in a series of MD simulations. We monitor the outer-shell radial fluctuation $R_{\mathrm{out}}(t)$, the angular-disorder parameter $D_{\theta}(t)$, and the Lindemann parameter $L(t)$. The vertical dashed line marks the first time at
which the melting criterion is satisfied, while the horizontal dashed line indicates the corresponding Lindemann threshold. The trajectories demonstrate
that the three observables do not evolve identically. Radial fluctuations typically increase early and may exhibit pronounced excursions well before the
melting time. The early growth of radial fluctuations observed here should not be identified with complete radial melting. Related simulations of finite two-dimensional Coulomb clusters have shown that localized radial rearrangements and subshell-specific radial disorder can occur before the loss of global radial shell order \cite{Tomecka2005}. Angular disorder grows more gradually, while the Lindemann parameter reaches its threshold only after the fluctuations have accumulated over an extended interval. A trajectory that does not melt may nevertheless display sizable transient radial excursions, showing that radial disorder is a precursor of melting rather than a sufficient condition for it.

\begin{figure}
    \centering
    \includegraphics[width=\columnwidth]{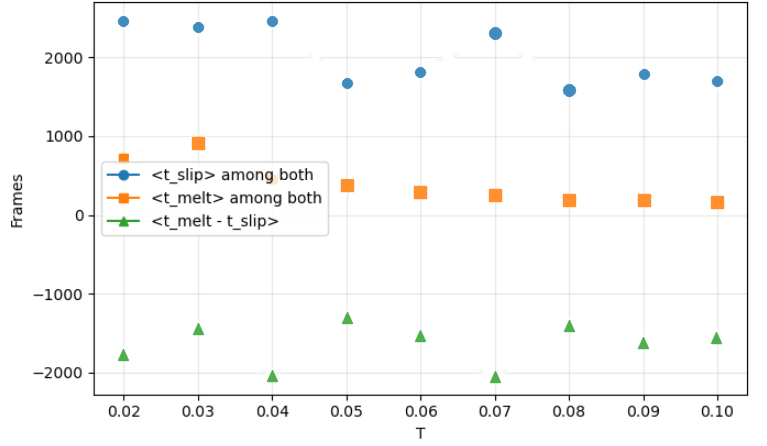}
    \caption{
    Mean event times for trajectories of the $(3,11)$ isomer in which both
    melting and large outer-shell motion occur as a function of temperature. Blue circles show the mean slip
    time, orange squares show the mean melting time, and green triangles show
    their difference. Melting systematically precedes the large collective
    shell motion.
    }
    \label{fig:event_timing}
\end{figure}

To quantify the temporal ordering of events statistically, we compared the mean onset time of large outer-shell motion with the mean melting time for trajectories in
which both events were detected. Figure~\ref{fig:event_timing} shows the result for the $(3,11)$ isomer of the $N=14$ crystal. Over the full temperature range, the characteristic time associated with large shell motion is substantially longer than the time at which the Lindemann melting criterion is reached.
Equivalently, the quantity $t_{\mathrm{melt}}-t_{\mathrm{slip}}$ remains negative. This result contradicts the hypothesis that melting is initiated by a completed azimuthal slip of the outer shell. Instead, the crystal satisfies the Lindemann-based dynamical melting criterion before a large collective shell rotation develops.

\begin{figure}
    \centering
    \includegraphics[width=0.48\textwidth]{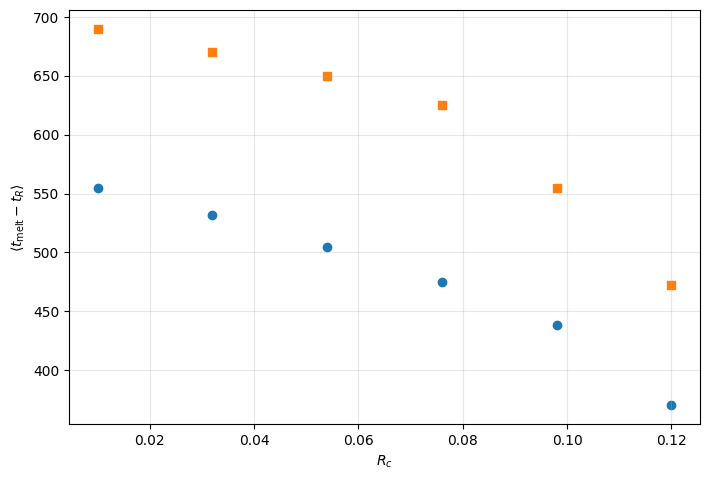}
    \caption{
    Mean time difference $\langle t_{\mathrm{melt}}-t_R\rangle$
    as a function of the radial-fluctuation threshold $R_c$ for trajectories
    of the $(3,11)$ isomer at $T=0.020$ and $T=0.040$.
    Circles and squares correspond to $T=0.020$ and $T=0.040$, respectively.
    The positive time difference over the entire investigated range of
    $R_c$ shows that the radial-fluctuation threshold is reached before
    melting for both temperatures, and that this ordering is insensitive
    to the precise choice of $R_c$.
    }
    \label{fig:threshold_robustness}
\end{figure}

The ordering of these events must not depend sensitively on the precise threshold used to define radial disorder. We therefore repeated the timing analysis for several values of the radial threshold $R_c$. The two temperatures
shown in Fig.~\ref{fig:threshold_robustness} give the same qualitative result: the mean difference between the melting time and the radial-fluctuation time is
positive over the entire investigated threshold interval. Thus, significant radial disorder develops before melting, and this ordering is robust against reasonable changes of the event definition.

\begin{figure}
    \centering
    \includegraphics[width=0.48\textwidth]{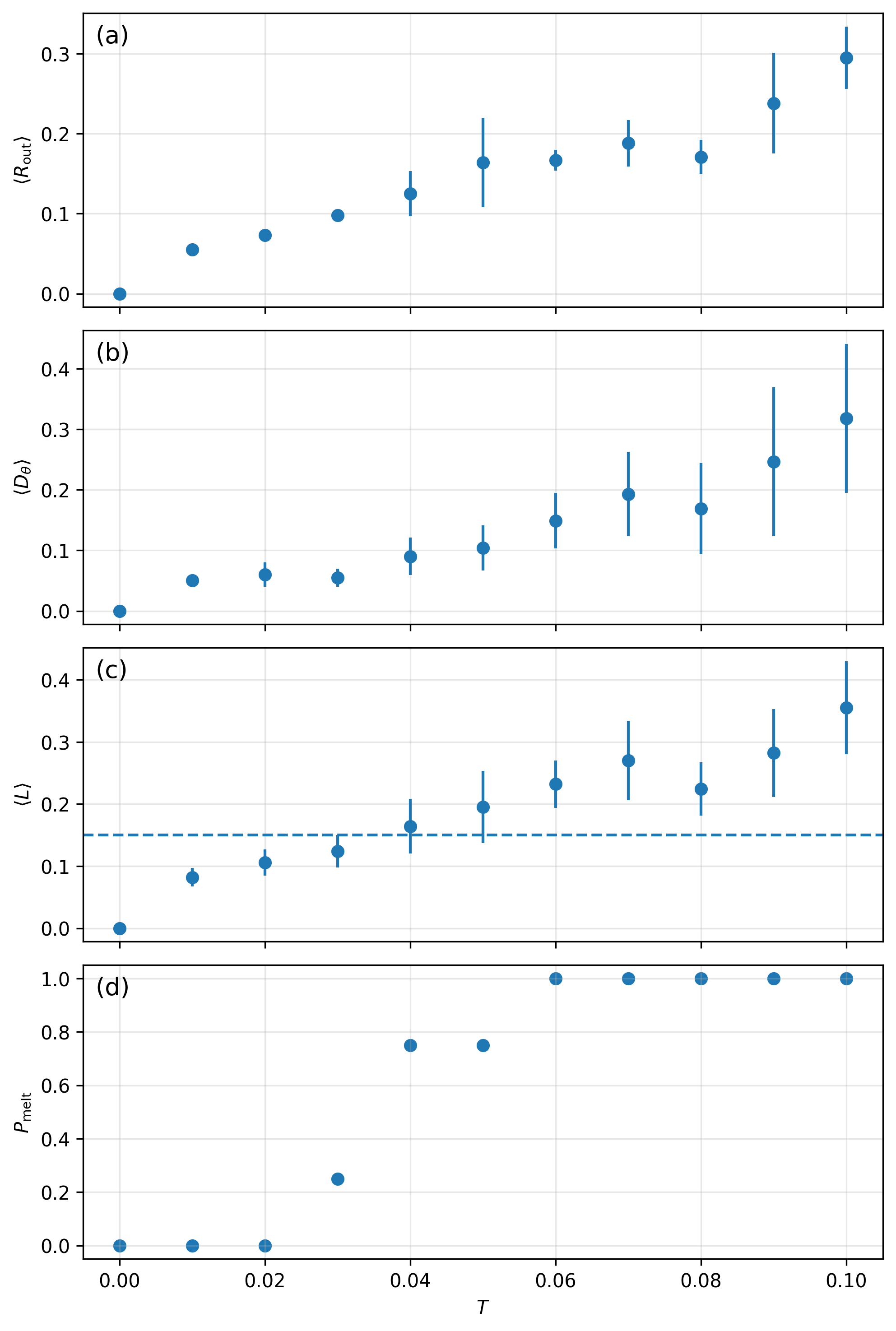}
    \caption{
    Temperature dependence of the ensemble-averaged observables characterizing
the melting dynamics and the corresponding melting probability:
(a) outer-shell radial fluctuation
$\langle R_{\mathrm{out}}\rangle$,
(b) angular disorder
$\langle D_{\theta}\rangle$,
(c) Lindemann parameter
$\langle L\rangle$, and
(d) melting probability $P_{\mathrm{melt}}$.
The dashed line in panel (c) marks the melting threshold $L=0.15$.
Radial fluctuations develop continuously with increasing temperature,
followed by increasing angular disorder and eventual loss of localization.
The melting probability increases over the same temperature interval in
which these observables exhibit their strongest growth.
Error bars in panels (a)--(c) represent the variation among independent
trajectories.
    }
    \label{fig:disorder_observables}
\end{figure}

The trajectory-level observations are consistent with the ensemble-averaged
temperature dependence shown in Fig.~\ref{fig:disorder_observables}. The mean
outer-shell radial fluctuation increases steadily with temperature and becomes
appreciable already below the main melting interval. The angular-disorder
parameter also increases, but its pronounced growth occurs at somewhat higher
temperatures and is accompanied by larger trajectory-to-trajectory
fluctuations. The Lindemann parameter follows the same overall trend and
crosses the melting threshold in the temperature range where the melting
probability rapidly rises. The conclusions presented below are insensitive to the particular choice of $L_{\mathrm{th}}$. Changing the threshold only changes the absolute value of the reported melting time, while preserving the relative ordering of the characteristic events observed during the melting process.

The corresponding melting probability is shown in
Fig.~\ref{fig:disorder_observables}. The transition occurs over the same
temperature range in which the Lindemann parameter crosses its threshold and both radial and angular disorder become substantial. The probability rises from approximately zero to unity over a finite temperature interval rather than at a sharply defined instability point. This behavior is consistent with a stochastic, thermally activated crossover in a finite system.

Taken together, these results do not support a picture in which melting is
caused by the sudden collapse of a single normal-mode frequency. No sharp
instability is observed in the fluctuation amplitudes. Instead, the radial,
angular, and Lindemann observables evolve smoothly with temperature. Such
behavior is compatible with progressive thermal broadening at approximately
fixed local stiffness, for which the fluctuation amplitude is expected to grow
approximately as
$\langle R^2\rangle\propto T/k_{\mathrm{eff}}$. A schematic summary of this melting scenario is provided in the Supplemental Information (Fig.~\ref{fig:melting_mechanism}).

The normal modes remain useful for identifying collective coordinates with large outer-shell participation. However, the transition is governed by the combined evolution of radial and tangential fluctuations rather than by motion
along a single soft coordinate. Radial disorder acts as an early precursor, angular disorder develops subsequently, and melting emerges only after the thermal fluctuations have accumulated sufficiently to destroy localization.

\subsection{Monte Carlo heat capacity}

\begin{figure}
    \centering
    \includegraphics[width=\columnwidth]{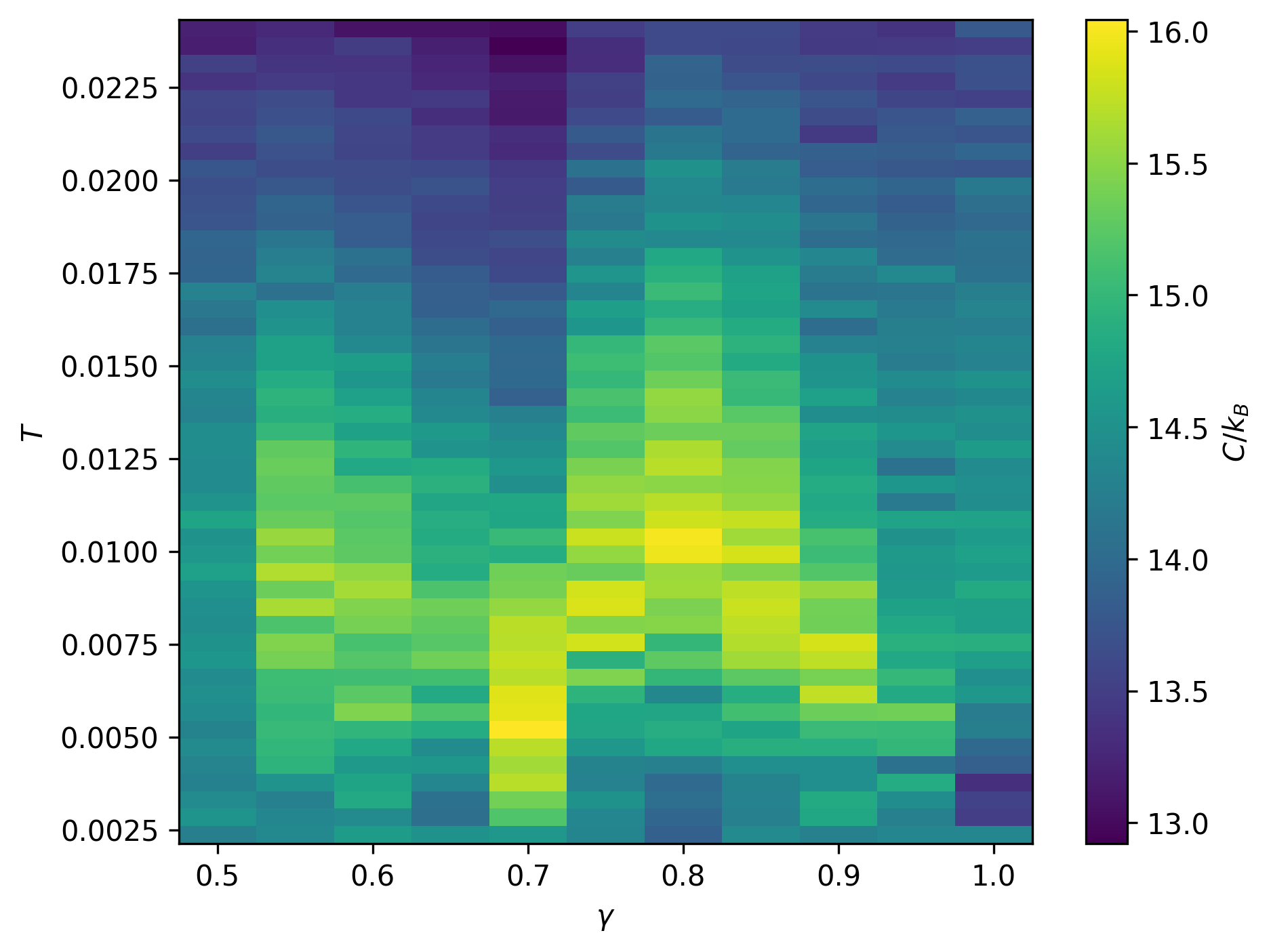}
    \caption{
Heat capacity obtained from Metropolis--Hastings Monte Carlo simulations as a function of temperature and trap anisotropy for a 14-ion crystal. The heat capacity was calculated from equilibrium energy fluctuations. The ridge of enhanced heat capacity indicates the temperature range where the crystal undergoes thermally induced disordering.
}
    \label{fig:mc_heatmap}
\end{figure}

Figure~\ref{fig:mc_heatmap} shows the heat capacity calculated using the Metropolis--Hastings algorithm as a function of temperature and trap anisotropy. A broad ridge of enhanced heat capacity is observed, indicating increased energy fluctuations associated with the structural disordering of the crystal.

The position of the heat-capacity maximum varies with trap anisotropy, following the same overall trend as the melting boundary obtained from molecular-dynamics simulations. This agreement supports the interpretation that the onset of melting is accompanied by enhanced equilibrium energy fluctuations.

The ridge occurs in approximately the same region of parameter space as the
melting crossover identified by molecular dynamics. Unlike molecular dynamics, the Monte Carlo approach does not provide information about transition pathways or nonequilibrium dynamics. Nevertheless, it offers an independent equilibrium characterization of the melting region through thermodynamic response functions.

\section{Discussion and Conclusions}

In this work we performed a systematic numerical investigation of melting in finite two-dimensional Coulomb crystals confined in an anisotropic harmonic trap. By combining energy minimization with molecular-dynamics simulations, we constructed melting phase diagrams for every crystal containing between $4\leq N\leq101$ ions. We observe that the melting transition becomes progressively sharper as the number of ions increases, with the measured transition width exhibiting an approximate $N^{-1/2}$ dependence over the investigated range. We also identify the important role played by structural isomerization. Finite Coulomb crystal isomers exhibit different thermal stability and therefore different melting probabilities and melting temperatures.

We performed a detailed analysis of the ion dynamics, which provides additional insight into the microscopic mechanism of melting. Contrary to the intuitive expectation that melting is triggered by excitation of the lowest-frequency tangential mode, our simulations show that radial fluctuations of the outer shell develop before the loss of localization and act as an early observable precursor of melting. 
Together with the smooth temperature dependence of the
radial, angular, and Lindemann observables, this behavior is consistent with
a gradual finite-system crossover rather than a sharp dynamical instability.
The shell-resolved dynamics is consistent with earlier observations of
inhomogeneous and orientational melting in confined Coulomb clusters
\cite{Bedanov1994,Kong2003,Apolinario2006Inhomogeneous,Lozovik1998}.

The isomer dependence is consistent with the sensitivity of finite-cluster
thermal behavior to metastable structure \cite{Yurtsever2005,Baletto2005Nanoclusters}. Recent studies have demonstrated configuration-resolved isomerization in two-dimensional ion crystals \cite{Mizukami2025} and thermally activated switching between trapped-ion configurations \cite{Ayyadevara2026ThermalActivation}. Our results extend this picture to melting, where the occupied isomer affects both the melting probability and pathway. Unlike the linear-to-zigzag transition \cite{Schiffer1993PhaseTransitions,Dubin1993StructuralTransitions,Fishman2008}, however, the observed melting is not associated with the softening of a single collective mode.

Although the calculations use an idealized classical model, the predicted
isomer-dependent stability and shell-resolved dynamics are directly accessible
in two-dimensional trapped-ion experiments. The phase diagrams provide
specific predictions for their dependence on crystal size and trap anisotropy.

An immediate extension is an investigation of larger crystals approaching the thermodynamic limit, where the evolution of the transition width and the emergence of
bulk-like behavior can be studied quantitatively. It will also be interesting to examine how the melting dynamics depends on the trap geometry, dimensionality,
and interaction potential, as well as to compare the predicted melting pathways with future experiments on two-dimensional trapped-ion crystals.
Such experimentally accessible predictions include the dependence of the melting temperature on crystal size and trap anisotropy, the shell-resolved character of melting, and the different thermal stability of competing structural isomers.

\bibliography{bib}
\clearpage
\appendix
\section*{Supplementary Information}
\label{sec:supplementary_information}

\subsection{Dimensionless formulation}

To reduce the number of independent parameters and facilitate the numerical analysis, all quantities are expressed in dimensionless units. Following Ref.~\cite{Pashinsky2025}, we introduce the characteristic length

\begin{equation}
l_0=
\left(
\frac{e^2}
{4\pi\varepsilon_0m\omega_y^2}
\right)^{1/3},
\end{equation}

\noindent obtained by equating the characteristic Coulomb interaction energy to the harmonic trapping energy. The corresponding characteristic energy and time are:

\begin{equation}
E_0=m\omega_y^2l_0^2,
\qquad
t_0=\omega_y^{-1}.
\end{equation}

\noindent Introducing the dimensionless coordinates $\tilde y_i={y_i}/{l_0}$
and
$\tilde z_i={z_i}/{l_0}$ and dividing Eq.~(\ref{eq:potential}) by the energy scale $E_0$, the total potential energy becomes

\begin{equation}
U=
\frac12
\sum_{i=1}^{N}
\left(
\tilde y_i^2
+
\gamma^2
\tilde z_i^2
\right)
+
\sum_{i<j}
\frac{1}
{\tilde r_{ij}}.
\label{eq:dimensionless_potential}
\end{equation}

In this representation all dimensional constants disappear from the equilibrium problem, leaving only the ion number $N$ and the trap anisotropy $\gamma$ as external control parameters. Throughout the remainder of this work the tildes are omitted, and all quantities are assumed to be expressed in these dimensionless units unless stated otherwise.

\subsection{Energy minimization}

For each pair of control parameters $(N,\gamma)$, equilibrium crystal configurations were obtained by minimizing the dimensionless potential energy given by Eq.~(\ref{eq:dimensionless_potential}). Since the potential-energy landscape of finite Coulomb crystals generally contains numerous local minima, especially for large ion numbers, the minimization procedure was repeated from many randomly generated initial configurations.

Each minimization converged to a mechanically stable equilibrium configuration corresponding to a local minimum of the potential energy. The remaining distinct minima were classified as structural isomers.

The global minimum was identified as the equilibrium crystal used throughout the melting calculations unless explicitly stated otherwise. Metastable isomers were retained for the analysis of structural isomerization presented in the \hyperref[sec:structural_isomerization]{``Structural isomerization" subsection}.

These equilibrium structures served as the initial conditions for all molecular-dynamics simulations described below.

\subsection{Molecular dynamics}
\label{sec:molecular_dynamics}

The thermal evolution of each equilibrium configuration was investigated using classical molecular-dynamics simulations. The equations of motion were numerically integrated
using a time step $\Delta t=10^{-2}$, for which convergence of the calculated melting observables with respect to the integration time step was verified. Unless stated otherwise, each trajectory contained $10^4$
integration steps.

Thermal initial conditions were generated by drawing the two velocity components of each ion independently from a Gaussian distribution. The velocities were then rescaled so that the total initial kinetic energy
satisfied
\begin{equation}
K_0=\frac{1}{2}NT ,
\end{equation}
where the ion mass is unity in the dimensionless units used here.
Independent trajectories therefore differed in their randomly generated initial velocities. After initialization, the system evolved under the harmonic confinement and Coulomb interaction without additional stochastic forcing or thermostating.

For the global melting phase diagrams, the melting probability at each $(N,\gamma,T)$ point was estimated adaptively. Sampling was initialized with a minimum ensemble size and additional trajectories were generated until the estimated binomial standard error fell below a prescribed threshold or the maximum ensemble size was reached.
\begin{equation}
\sigma_P =
\sqrt{\frac{P_{\rm melt}(1-P_{\rm melt})}{N_{\rm traj}}}
\end{equation}
Thus, the statistical uncertainty decreases as $N_{\rm traj}^{-1/2}$ and is largest near $P_{\rm melt}=1/2$, and
fell below $0.01$, or until a maximum of $500$ trajectories was reached. The resulting ensemble size $N_{\rm traj}$ therefore depended on the melting probability at the corresponding point.

\subsection{Melting criterion}
\label{sec:melting_criterion}

Since finite Coulomb crystals do not exhibit a true thermodynamic phase transition, melting must be identified through an operational criterion applied to individual molecular-dynamics trajectories. In this work, melting is defined as the loss of localization of at least one ion relative to its equilibrium position.

For each equilibrium configuration, we first compute the nearest-neighbor distance scale

\begin{equation}
d_{\min}
=
\min_i
\left[
\min_{j\neq i}
\left|
\mathbf r_i^{(0)}-\mathbf r_j^{(0)}
\right|
\right],
\end{equation}

\noindent where $\mathbf r_i^{(0)}$ is the equilibrium position of ion $i$. During the molecular-dynamics simulation, the instantaneous displacement of each ion from its initial equilibrium position is monitored,

\begin{equation}
\Delta_i(t)
=
\left|
\mathbf r_i(t)-\mathbf r_i^{(0)}
\right|.
\end{equation}

\noindent A trajectory is classified as melted once

\begin{equation}
\max_i \Delta_i(t) > 2 d_{\min}.
\label{eq:melting_criterion}
\end{equation}

This criterion identifies the point at which at least one ion becomes delocalized by more than twice the shortest equilibrium nearest-neighbor spacing. It is therefore sensitive to large-amplitude motion and shell delocalization rather than to small thermal vibrations around equilibrium.

Because melting in a finite crystal is a stochastic thermally activated process, a single trajectory cannot define a sharp melting boundary. For each combination of ion number $N$, trap anisotropy $\gamma$, and temperature $T$, we therefore generate an ensemble of independent molecular-dynamics trajectories. Each trajectory starts from the same equilibrium configuration but uses an independent realization of the thermal noise.

If number of trajectories $n_{\rm melt}$ out of the total number $N_{\rm traj}$ trajectories satisfy Eq.~(\ref{eq:melting_criterion}), then the melting probability is estimated as

\begin{equation}
P_{\rm melt}
=
\frac{n_{\rm melt}}{N_{\rm traj}}.
\label{eq:melting_probability}
\end{equation}

The melting phase diagrams shown below are obtained by evaluating $P_{\rm melt}$ on a grid of temperatures and trap anisotropies. Each point in such a diagram therefore represents an ensemble-averaged probability rather than the outcome of a single simulation.

\subsection{Linear stability and normal modes}

To characterize the collective motion of the crystal near equilibrium, we
performed a normal-mode analysis for each mechanically stable configuration.
The ion positions were written as small displacements $\delta \mathbf r_i(t)$ around the equilibrium,

\begin{equation}
\mathbf r_i(t)
=
\mathbf r_i^{(0)}
+
\delta \mathbf r_i(t).
\end{equation}

\noindent Then, expanding the dimensionless potential energy to second order gives

\begin{equation}
U
=
U_0
+
\frac12
\delta \mathbf R^{T}
H
\delta \mathbf R
+
O(\delta R^3),
\end{equation}

\noindent where

\begin{equation}
\delta \mathbf R
=
(\delta y_1,\ldots,\delta y_N,
\delta z_1,\ldots,\delta z_N)^{T},
\end{equation}

\noindent and $H$ is the Hessian matrix,

\begin{equation}
H_{\mu\nu}
=
\left.
\frac{\partial^2 U}
{\partial R_\mu \partial R_\nu}
\right|_{\mathbf R=\mathbf R^{(0)}} .
\end{equation}

\noindent The normal modes were obtained by diagonalizing the Hessian,

\begin{equation}
H\mathbf e_k
=
\omega_k^2
\mathbf e_k ,
\end{equation}

\noindent where $\omega_k$ is the dimensionless frequency of mode $k$ and
$\mathbf e_k$ is the corresponding normalized eigenvector.

To quantify the spatial and directional character of each normal mode, we introduce two additional measures. For a normalized eigenmode
$\mathbf e_k$, let $\mathbf v_i^{(k)}$ denote the two-dimensional displacement vector of ion $i$ in that mode.
Then the outer-shell participation is defined as

\begin{equation}
P_{\mathrm{out}}^{(k)}
=
\frac{
\displaystyle
\sum_{i\in\mathcal S_{\mathrm{out}}}
\left|\mathbf v_i^{(k)}\right|^2
}{
\displaystyle
\sum_{i=1}^{N}
\left|\mathbf v_i^{(k)}\right|^2
},
\label{eq:outer_participation}
\end{equation}

\noindent where $\mathcal S_{\mathrm{out}}$ denotes the set of ions belonging to the outer shell. Thus, $P_{\mathrm{out}}^{(k)}=1$ corresponds to a mode completely
localized on the outer shell.

To distinguish the radial and the tangential motion, a local unit tangent
$\mathbf t_i$ is constructed along the elliptical equilibrium shell in the stretched coordinate system $(y,\gamma z)$ defined in the next subsection. The tangential fraction of mode $k$ is then defined as

\begin{equation}
f_{\theta}^{(k)}
=
\frac{
\displaystyle
\sum_{i\in\mathcal S_{\mathrm{out}}}
\left(
\mathbf v_i^{(k)}\cdot\mathbf t_i
\right)^2
}{
\displaystyle
\sum_{i=1}^{N}
\left|\mathbf v_i^{(k)}\right|^2
}.
\label{eq:tangential_fraction}
\end{equation}

Thus, $P_{\mathrm{out}}^{(k)}$ measures how strongly the mode is localized on the outer shell, whereas $f_{\theta}^{(k)}$ measures the fraction of the total mode amplitude corresponding specifically to tangential motion of the outer-shell ions.

\subsection{Dynamical observables}
\label{sec:dynamical_observables}

To characterize the microscopic evolution of the crystal during melting, we
analyze several trajectory-dependent observables describing radial motion,
angular motion, collective outer-shell rotation, and the overall loss of
particle localization. All quantities are calculated for every stored
molecular-dynamics frame. When temperature-dependent results are reported, the
corresponding quantities are subsequently averaged over independent
trajectories.

Because the equilibrium shells are elongated in an anisotropic trap, radial
and angular coordinates are defined in the stretched coordinate system
$(y,\gamma z)$. For ion $i$, we introduce

\begin{equation}
    \rho_i(t)
    =
    \sqrt{y_i^2(t)+\gamma^2 z_i^2(t)}
\end{equation}

and

\begin{equation}
\theta_i(t)
=
\arg\!\left(y_i+i\gamma z_i\right),
\end{equation}

The set of ions belonging to the outer shell,
$\mathcal{S}_{\mathrm{out}}$, is determined from the equilibrium configuration
using the shell-identification procedure described above. The same shell
assignment is then retained throughout the trajectory analysis.

The radial deformation of the outer shell is characterized by

\begin{equation}
    R_{\mathrm{out}}(t)
    =
    \left[
    \frac{1}{N_{\mathrm{out}}}
    \sum_{i\in\mathcal{S}_{\mathrm{out}}}
    \left(
        \rho_i(t)-\rho_i^{(0)}
    \right)^2
    \right]^{1/2},
    \label{eq:Rout}
\end{equation}

\noindent where $\rho_i^{(0)}$ is the equilibrium elliptic radius of ion $i$ and $N_{\mathrm{out}}$ is the number of ions in the outer shell. This quantity measures radial fluctuations of the shell while remaining insensitive to a purely rigid azimuthal rotation.

Angular motion of the outer shell is characterized by the root-mean-square
angular displacement

\begin{equation}
    D_{\theta}(t)
    =
    \left[
    \frac{1}{N_{\mathrm{out}}}
    \sum_{i\in\mathcal{S}_{\mathrm{out}}}
    \Delta\theta_i^2(t)
    \right]^{1/2},
    \label{eq:Dtheta}
\end{equation}

\noindent where

\begin{equation}
    \Delta\theta_i(t)
    =
    \operatorname{Arg}
    \left\{
    \exp\left[
        i\left(
        \theta_i(t)-\theta_i^{(0)}
        \right)
    \right]
    \right\}
\end{equation}

\noindent is the wrapped angular displacement of ion $i$ from its equilibrium angle.
The wrapping removes artificial discontinuities at $\pm\pi$.
The same definition of $D_\theta$ is used for the individual time traces and
for the ensemble-averaged temperature dependence shown below. Since
$D_\theta$ is measured relative to the equilibrium angles, it includes both
coherent rotation of the outer shell and non-rigid angular rearrangements.

The overall loss of particle localization is monitored using the Lindemann
parameter. For a trajectory extending from the initial frame to time $t$, it is
defined as

\begin{equation}
    L(t)
    =
    \frac{1}{\bar d_{\mathrm{nn}}}
    \left[
    \frac{1}{N}
    \sum_{i=1}^{N}
    \frac{1}{N_t}
    \sum_{\tau=0}^{t}
    \left|
        \mathbf r_i(\tau)-\mathbf r_i(0)
    \right|^2
    \right]^{1/2},
    \label{eq:lindemann_dynamic}
\end{equation}

\noindent where $\bar d_{\mathrm{nn}}$ is the mean equilibrium nearest-neighbor distance
and $N_t$ is the number of stored frames included in the average. Thus,
$L(t)$ is a cumulative measure of particle delocalization rather than an
instantaneous displacement.

For the trajectory-resolved dynamical analysis, the melting time is defined as
the first threshold-crossing time

\begin{equation}
    t_{\mathrm{melt}}
    =
    \min
    \left\{
        t:L(t)\geq L_{\mathrm{th}}
    \right\},
    \qquad
    L_{\mathrm{th}}=0.15.
    \label{eq:tmelt}
\end{equation}

Here, the value $L_{\mathrm{th}}=0.15$ is used as an operational marker for comparing different trajectories. It is not interpreted as a universal critical value. Changing the threshold modifies the absolute crossing time but does not affect the qualitative ordering of the dynamical events discussed below.

This time-resolved definition is distinct from the displacement criterion used to classify complete trajectories in the global melting phase diagrams. The displacement criterion provides a binary melted or non-melted classification, whereas the Lindemann threshold provides a convenient continuous marker of the melting time within an individual trajectory.

To determine whether radial fluctuations precede melting, we introduce a radial threshold $R_c$ and define

\begin{equation}
    t_R(R_c)
    =
    \min
    \left\{
        t:R_{\mathrm{out}}(t)\geq R_c
    \right\}.
    \label{eq:tR}
\end{equation}

The comparison between $t_R$ and $t_{\mathrm{melt}}$ is repeated for a range of $R_c$ values. This procedure tests whether the temporal ordering of radial fluctuations and melting is robust against the particular operational choice of the radial threshold.

Large collective azimuthal motion of the outer shell is analyzed separately from the total angular displacement $D_\theta$. For each frame, we determine the rigid angular shift $\phi(t)$ that best aligns the instantaneous
outer-shell angles with their equilibrium values. The residual displacement of ion $i$ after removal of this rigid rotation is

\begin{equation}
    \delta\theta_i(t)
    =
    \operatorname{wrap}
    \left[
        \theta_i(t)-\theta_i^{(0)}-\phi(t)
    \right].
\end{equation}
where $\operatorname{wrap}(x)$ maps the angular difference onto the principal interval $(-\pi,\pi]$. A collective slip is identified when the unwrapped rigid rotation reaches half
of the equilibrium angular spacing between neighboring outer-shell ions,
\begin{equation}
    \left|\phi(t)\right|
    \geq
    \phi_{\mathrm{slip}},
    \qquad
    \phi_{\mathrm{slip}}
    =
    \frac{\pi}{N_{\mathrm{out}}},
    \label{eq:slip_threshold}
\end{equation}
while the residual deformation remains sufficiently small for the motion to be
classified as approximately rigid. The slip time is then defined as

\begin{equation}
    t_{\mathrm{slip}}
    =
    \min
    \left\{
        t:
        |\phi(t)|\geq\phi_{\mathrm{slip}}
    \right\}.
    \label{eq:tslip}
\end{equation}

In comparisons of characteristic event times, averages are taken only over
trajectories in which all relevant events occur within the simulated time
window. For example, the comparison of $t_{\mathrm{melt}}$ and
$t_{\mathrm{slip}}$ includes only trajectories in which both events are
detected. Trajectories without a detected event remain included in
ensemble-averaged observables and melting probabilities, but are excluded from
conditional averages of the corresponding event time.

\subsection{Metropolis Monte Carlo}

As an independent equilibrium cross-check of the molecular-dynamics melting region, we also performed equilibrium Monte Carlo simulations using the Metropolis--Hastings algorithm ~\cite{Metropolis1953,Hastings1970}. Unlike molecular dynamics, which follows deterministic trajectories governed by Newton's equations, the Monte Carlo approach samples equilibrium configurations directly from the canonical Gibbs distribution.

Starting from an equilibrium crystal configuration, trial particle displacements were generated randomly and accepted with probability

\begin{equation}
P_{\mathrm{acc}}=
\min\left[
1,
\exp\left(
-\frac{\Delta U}{k_B T}
\right)
\right],
\end{equation}

\noindent where $\Delta U$ is the change in the total potential energy resulting from the proposed move. After equilibration, the sampled configurations were used to evaluate equilibrium averages of the potential energy and its fluctuations.

The heat capacity $C$ was obtained from the fluctuation relation

\begin{equation}
C=
\frac{\langle E^2\rangle-\langle E\rangle^2}
{k_B T^2},
\end{equation}

where the brackets denote averages over the Monte Carlo ensemble. This approach allows thermodynamic quantities to be determined without integrating the equations of motion and therefore provides an independent estimate of the melting region.

\subsection{Schematic picture of the melting dynamics}

\begin{figure*}
    \centering
    \includegraphics[width=\textwidth]{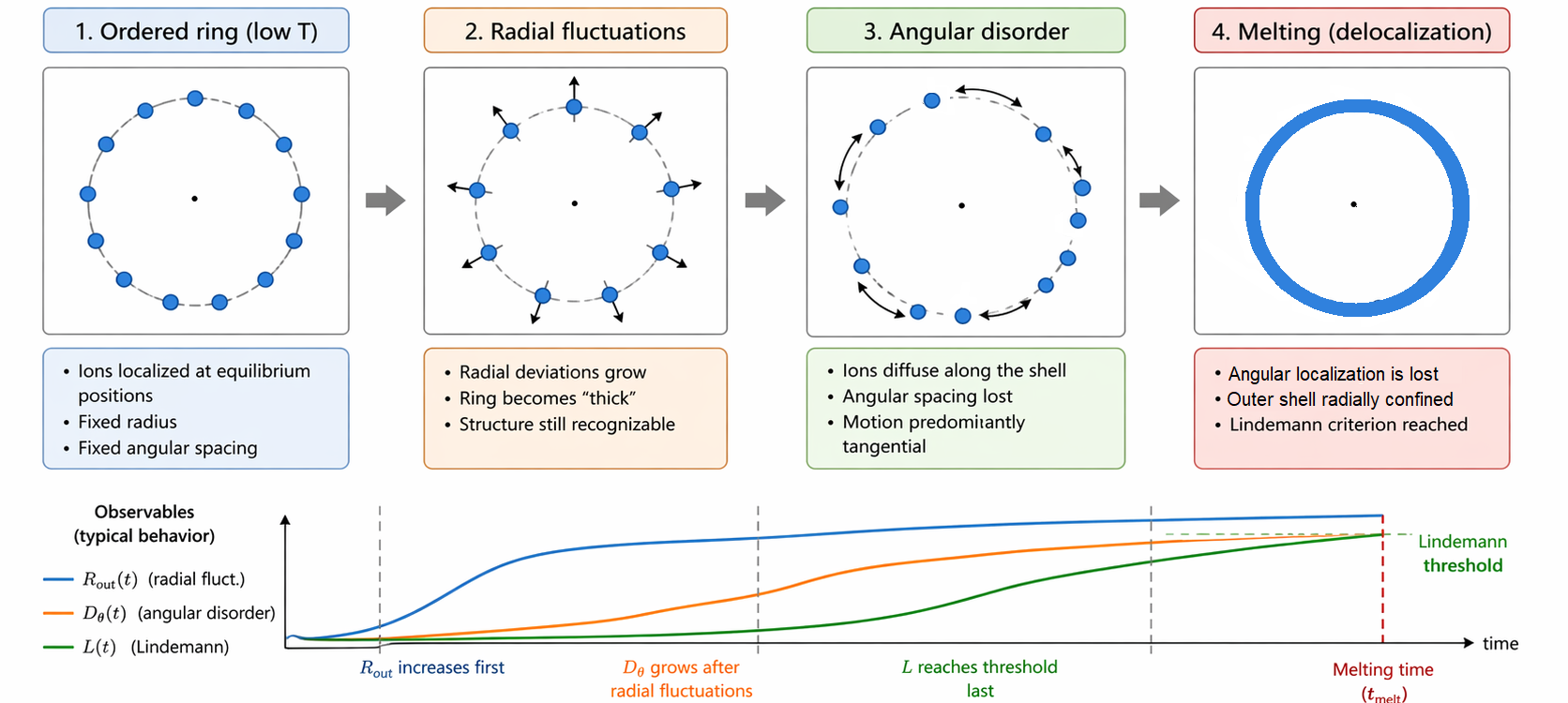}
    \caption{
    Schematic microscopic mechanism of melting in a finite two-dimensional ion crystal. Thermal fluctuations first enhance radial excursions of the outer-shell ions. The radial distortion reduces the constraints on motion along the shell, allowing increasingly large tangential displacements. The resulting loss of angular localization produces the melted outer-shell
    state. The observed tangential motion is therefore a consequence and characteristic feature of melting rather than its earliest precursor.
    }
    \label{fig:melting_mechanism}
\end{figure*}

Figure~\ref{fig:melting_mechanism} summarizes the microscopic picture
of the melting dynamics inferred from the trajectory analysis.
Thermal excitation first produces radial fluctuations of the outer shell.
These excursions weaken the geometric constraints imposed by neighboring
ions and open additional pathways in configuration space. The ions then
acquire greater freedom to move tangentially along the shell. Angular localization is progressively lost, and the motion becomes increasingly
azimuthal in the melted state.

Thus, tangential motion is a prominent signature of the melted phase,
but it is not the initial trigger of melting. The results support a picture
in which radial disorder acts as an early precursor of melting, whereas
pronounced tangential motion develops predominantly after localization
has already been lost. Shell-dependent and orientational melting have
previously been reported in classical confined clusters and experimentally
observed ion crystals
\cite{Apolinario2006Inhomogeneous,Duca2023,Pashinsky2025}.

\end{document}